\documentclass[10pt]{iopart}

\usepackage{iopams,cite}  
\usepackage{graphicx}
\usepackage{dcolumn}
\usepackage{color}
\usepackage[displaymath,mathlines]{lineno}
\begin{document}

\title[Breaking the ring: $^{53}$Cr-NMR on the Cr$_{8}$Cd molecular nanomagnet]{Breaking the ring: $^{53}$Cr-NMR on the Cr$_{8}$Cd molecular nanomagnet}

\author{E. Garlatti$^1$, G. Allodi$^1$, S. Bordignon$^1$, L. Bordonali$^2$, G.A. Timco$^3$, R.E.P. Winpenny$^3$, A. Lascialfari$^4$, R. De Renzi$^1$ and S. Carretta$^{1,5}$}

\address{$^1$ Dipartimento di Science Matematiche, Fisiche e Informatiche,
	Universit\`{a} di Parma, Parco Area delle Scienze 7/A, 43124 Parma, Italy.}
\address{$^2$ Karlsruhe Institute of Technology,
	Institute of Microstructure Technology, Hermann-von-Helmholtz-Platz 1 76344 Eggenstein-Leopoldshafen, Germany.}
\address{$^3$ School of Chemistry and Photon Institute, The University of Manchester, M13 9PL Manchester, United Kingdom.}
\address{$^4$ Dipartimento di Fisica, Universit\`{a} degli Studi di Pavia, Via Bassi 6, 27100 Pavia, Italy and INFN, Milano Unit, Milano, Italy.}
\address{$^5$ INSTM, UdR Parma, 43124 Parma, Italy.}

\ead{stefano.carretta@unipr.it}

\vspace{10pt}
\begin{indented}
\item[]February 2020
\end{indented}

\begin{abstract}
An accurate experimental characterization of finite antiferromagnetic (AF) spin chains is crucial for controlling and manipulating their magnetic properties and quantum states for potential applications in spintronics or quantum computation. In particular, finite AF chains are expected to show a different magnetic behaviour depending on their length and topology. Molecular AF rings are able to combine the quantum-magnetic behaviour of AF chains with a very remarkable tunability of their topological and geometrical properties. In this work we measure the $^{53}$Cr-NMR spectra of the Cr$_8$Cd ring to study the local spin densities on the Cr sites. Cr$_8$Cd can in fact be considered a model system of a finite AF open chain with an even number of spins. The NMR resonant frequencies are in good agreement with the theoretical local spin densities, by assuming a core polarization field $A_C$ = -12.7 T/$\mu_B$. Moreover, these NMR results confirm the theoretically predicted non-collinear spin arrangement along the Cr$_8$Cd ring, which is typical of an even-open AF spin chain.

\end{abstract}

%
%
%
\ioptwocol

\section{Introduction}

Quantum properties of matter have been in the spotlight of fundamental research in physics for decades, but nowadays they also represent essential and solid resources for the future of nanoscience and nanotechnology. In particular, magnetic properties of quantum systems are of central interests for the current research in condensed matter physics. 
Topological effects  have a strong influence on low-dimensional magnetic systems and on the excitation spectrum and eigenfunctions of antiferromagnetic (AF) Heisenberg spin chains \cite{due,tre}. The experimental observation of the so-called ``edge states'' in broken $S = 1$ AF Heisenberg spin chains \cite{edge1} also triggered the interest in finite open (or ``broken'') chains, making them the subject of several theoretical investigations in the past decades. These works showed that the magnetic behaviour of finite open AF chains depends on their length, topology (closed vs open) and parity of the number of atoms (even vs odd) \cite{dodici,edge3,undici,Politi,tredici}. For instance, finite even-open chains are predicted to show a Non-Collinear (NC) configuration of the local-spin $s_{i}$ expectation values for $s_{i} > 1/2$ \cite{edge3}. Conversely, odd-open chains below a critical length are arranged in a ferrimagnetic configuration \cite{undici}, while even-closed ones in a spin-flop arrangement \cite{Politi}. It is also worth to stress that all these topology and parity effects are already evident for very short chains with a number of sites $N \leq 10$.\\
Understanding the magnetic behaviour of finite AF spin chains is not only of fundamental interest for quantum magnetism, but it is also interesting for potential applications in spintronics or quantum computation \cite{otto,nove,dieci}. Thus, an accurate experimental characterization of these systems is crucial for controlling their magnetism and manipulating their quantum states. Molecular nanomagnets \cite{Sessoli1993} and in particular AF rings are able to combine the quantum-magnetic behaviour of finite AF chains with a very remarkable tunability of their topological and geometrical properties, offering several experimental advantages. Thanks to the high degree of control achieved in the chemistry of magnetic molecules, it is in fact possible to synthesize rings with an even or odd number of magnetic ions with integer or half-integer spins. For instance, $s=3/2$ homometallic rings like Cr$_8$ and Cr$_9$ \cite{Carretta2003, Affronte2003, Waldmann2003,NatPhys,Baker2012,Waldmann2009,Ummethum2012,Cu3,Cr8Ni1,Cr8Ni2,VO7,Antkowiak2013,Kamieniarz2015,Garlatti2016} allow the study of closed chains, thanks to their periodic boundary conditions, and to investigate how frustration influence the magnetic behaviour of the ring/chain. The introduction of non-magnetic impurities breaking the cyclic symmetry as in Cr$_7$Cd \cite{Micotti2006,Piligkos2007} and Cr$_8$Cd or Cr$_8$Zn \cite{Timco2005,Furukava2008,Bianchi2009,Adelnia2015} makes these rings effective model systems for open chains.\\
Heterometallic rings \cite{Larsen2003,Guidi2005,Garlatti2014b,Cr7Co} with a magnetic ground state can be obtained by chemical substitution of one or two magnetic centers. Among them, Cr$_7$Ni has been proposed as a qubit for quantum computation \cite{Troiani20051,Troiani20052,NatNano,Wedge2012,Ferrando2016,Garlatti2017}. Indeed, the investigation of the local spin structure of AF rings not only provides a quantitative insight on the behaviour of AF spin chains, but it is also very important in the design of supramolecular chains for quantum information processing. For instance, these molecular rings can be linked together, either directly or through magnetic ions, into supramolecular interacting dimers \cite{NatNano,Ferrando2016,Garlatti2017}. The site dependence of the local spin density plays a key role in the scheme proposed for obtaining time dependent qubit-qubit couplings in the presence of permanent exchange interactions \cite{Wedge2012}. Thus, it is crucial to understand how the local spin density depends on the length and topology of the investigated ring.\\  
The local spin density along the odd-membered open ring Cr$_{7}$Cd has been previously determined through $^{53}$Cr-NMR \cite{Micotti2006}. That work showed that the local spin density in the ground state is rather uniformly distributed over the ring with an alternated staggered orientation due to the AF coupling. Then, we investigated how the single-ion spin moment is distributed in an heterometallic ring, where one Cr$^{3+}$ ion is replaced by a different magnetic ion rather than a diamagnetic one \cite{JPCM2012}. The $^{53}$Cr-NMR spectra measured at low temperature in a single crystal of the Cr$_7$Ni in its $S=1/2$ ground state are in agreement with the theoretically predicted AF staggered configuration, with the Cr$^{3+}$ ions next to the Ni$^{2+}$ ion displaying the greatest component of the local spin $s_{i}$.\\
The spin arrangement for even-open rings like Cr$_{8}$Cd is expected to be different both from the spin-flop configuration of even-closed rings, like Cr$_{8}$, and from the AF staggered arrangement of odd-open ones, like Cr$_{7}$Cd. Theoretical models in fact predict a NC configuration \cite{edge3,NatCommun2015}. In this configuration, the spins at the extremities of the chain have the highest local magnetic moment along the direction of the applied magnetic field, since they have to compete only with one exchange interaction. Moving towards the centre of the chain, nearest-neighbouring spins are aligned in the opposite direction due to AF interactions and the magnetic moment along the applied field decreases. For the two spins at the exact centre of the even chain, the AF interactions between them and their nearest-neighbours cannot be simultaneously satisfied, leading to competing interactions.\\
In this work we measure the $^{53}$Cr-NMR spectra of Cr$_8$Cd to study the local spin densities on the Cr sites, as a model system of a finite Heisenberg AF even-open chain. Cr$_8$Cd has already been studied by polarized neutron diffraction (PND) \cite{NatCommun2015}. PND allowed to extract the thermal averages of the projected local spin moments along the applied magnetic field, which were found to be in agreement with the theoretically predicted NC configuration. The same technique also confirmed the spin-flop configuration of Cr$_8$ \cite{NatCommun2015}. Here we measure the even-open ring Cr$_8$Cd with $^{53}$Cr-NMR to compare it with previous NMR results on other AF rings and to confirm polarised neutron results. Moreover, with $^{53}$Cr-NMR we are able to obtain the expectation values of local spin density along the ring, together with the core polarisation field of $^{53}$Cr nuclei. Our experimental results are in agreement with theoretical predictions, confirming the NC spin configuration for Cr$_8$Cd. It is important to note that NMR is a less resource- and time-consuming technique with respect to PND and allows one to determine local spin moments after a straightforward data analysis. NMR experiments similar to the one presented here on Cr$_8$Cd can in fact be devised for more complex AF rings or long AF chains, allowing one to obtain local spin moments with the level of accuracy required for quantum computation applications. On the contrary, the complexity of the interpretation of PND data increases with the complexity of the investigated system. For instance, PND data analysis requires a very accurate determination of the molecular structure of sample from neutron or X-rays diffraction experiments as a starting point, on which the outcome of the analysis depends critically. 


\section{Experimental results}

\begin{figure}[!h]
	\centering
	\includegraphics[width=0.5\textwidth]{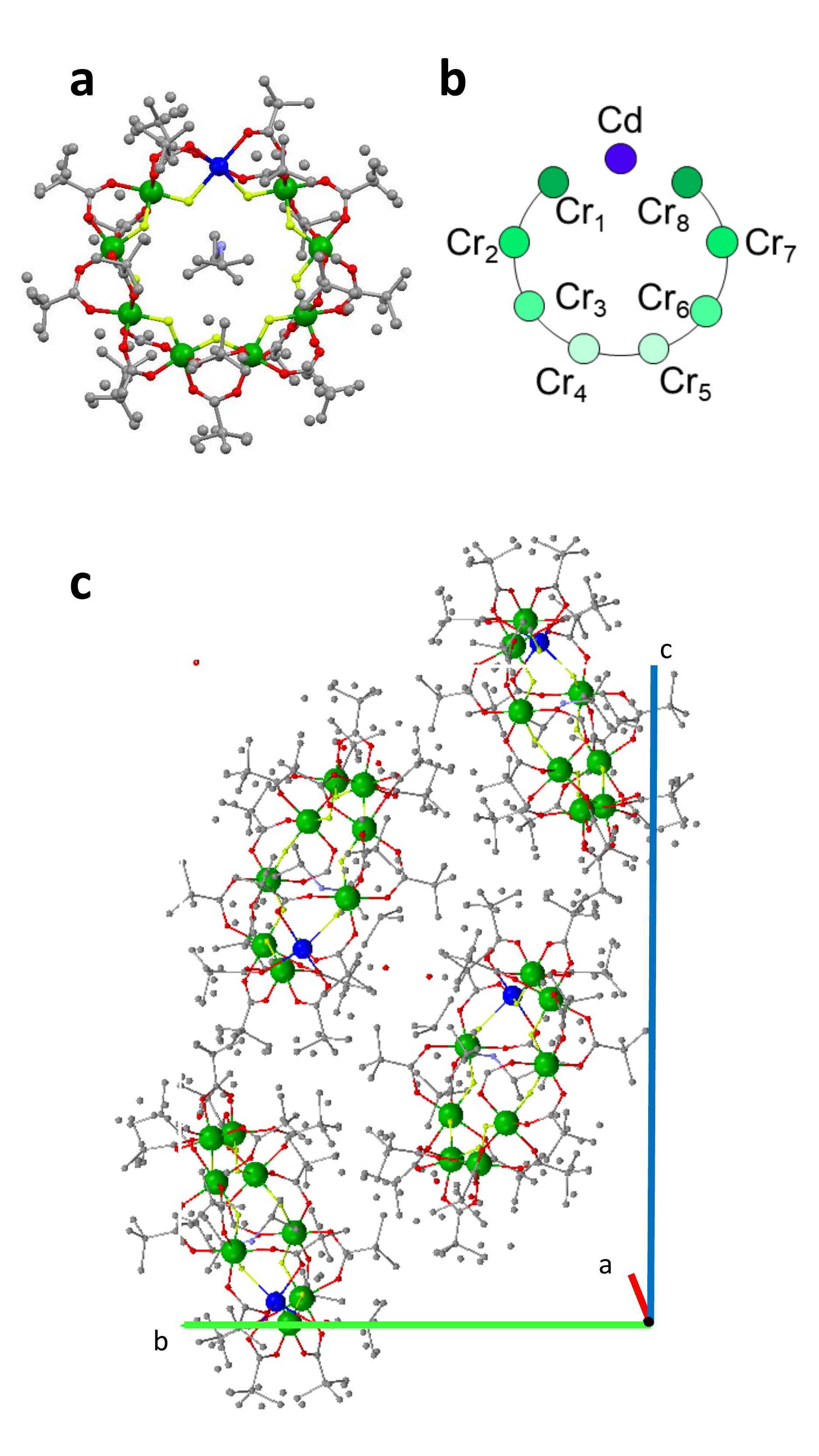}
	\caption{Panel a: [H$_2$N$^t$Bu$^{is}$Pr][Cr$_8$CdF$_9$(O$_2$CCMe$_3$)$_{18}$] molecular structure. Cr ions are reported in green, Cd in blue, O in red, F in yellow and C in grey. Hydrogen ions are omitted for clarity. Panel b: schematic representation of ions along the ring, the color-code highlights the magnetically equivalent Cr$^{^3+}$ ions and the non-magnetic Cd$^{2+}$. Panel c: two inequivalent molecules are clearly distinguishable within the unit cell.}
	\label{molecule}
\end{figure}

[H$_2$N$^t$Bu$^{is}$Pr][Cr$_8$CdF$_9$(O$_2$CCMe$_3$)$_{18}$], Cr$_8$Cd in short, is a heterometallic AF ring constituted of nine transition metal ions (see Fig.\ref{molecule}-a), obtained from its parent compound Cr$_8$ by adding a $s=0$ Cd$^{2+}$ ion to the eight Cr$^{3+}$. We performed our NMR experiment on a single crystal sample, prepared as in Ref.\cite{Timco2005}. The unit cell contains two identical but differently oriented molecules (see Fig.\ref{molecule}-c), each making an angle of 27$^{\circ}$ between the perpendicular to the plane of the ring (i.e., $z$ axis of the molecule) and the b axis. The static magnetic field has therefore been applied along the b axis of the crystal, in order to ensures equal components of the field along the $z$-axis for all the molecules in the unit cell.\\
The main difficulty of $^{53}$Cr-NMR measurements is due to the low natural abundance and low sensitivity of the probe (9.54 $\%$ abundance, $\gamma/2\pi$ = 2.406 MHzT$^{-1}$), together with the millimetric size of the measured crystal. Despite this drawback, we managed to observe signals at low temperature in the desired magnetic field range. The experiments have been performed in a Maglab EXA$^{\circledR}$ (Oxford Instruments) cold-bore field-sweeping superconducting cryomagnet, with a variable temperature insert (VTI) as a sample environment, and a ``HyReSpect'' home-built NMR spectrometer \cite{Hyrespect}, equipped with an external rf power amplifier. The magnet allows to cover a of 0-9 T static magnetic field range, while the VTI allows us to reach a base temperature of 1.4 K, at which all measurements have been performed.\\
\begin{figure}[!h]
	\centering
	\includegraphics[width=0.48\textwidth]{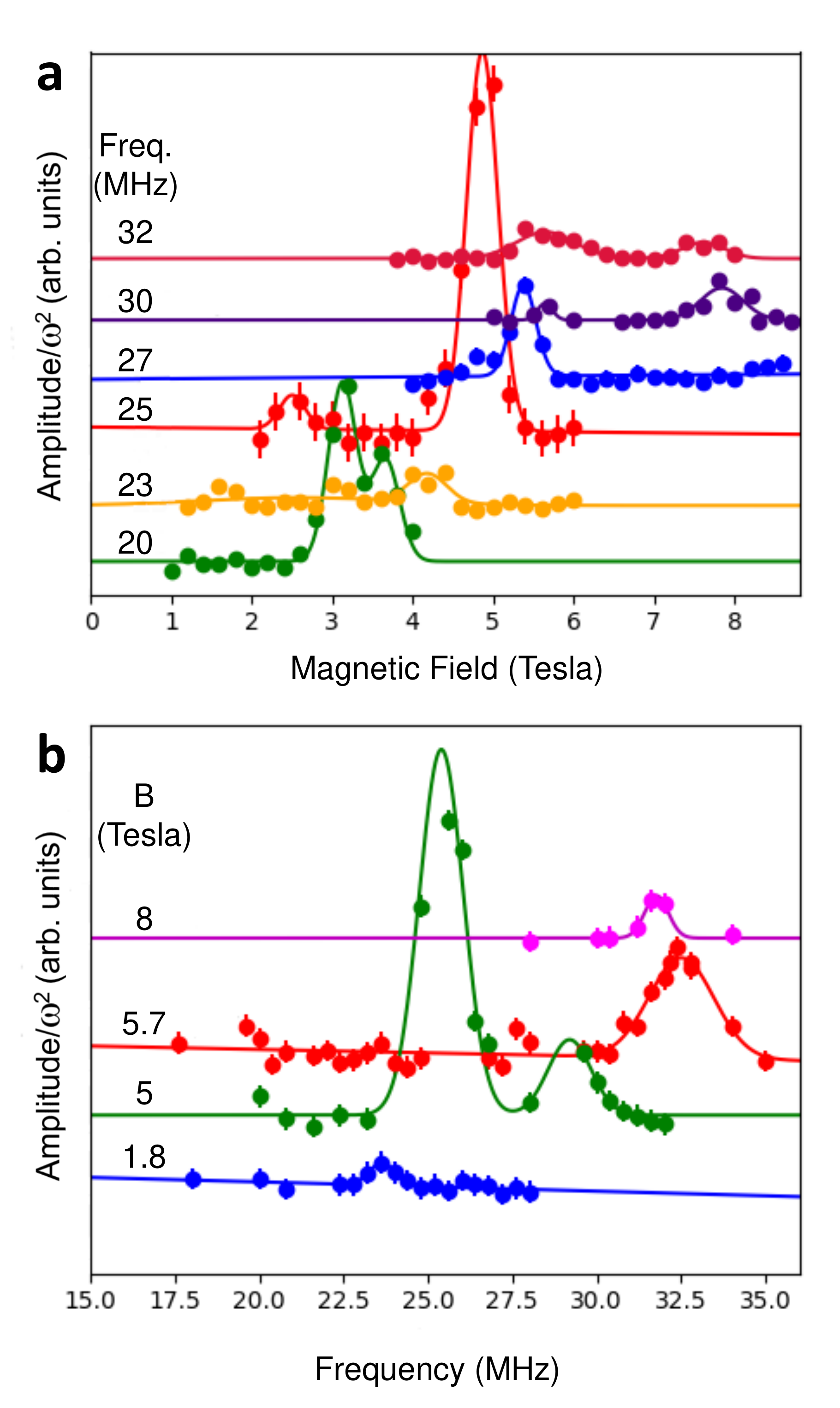}
	\caption{$^{53}$Cr-NMR spectra of Cr$_8$Cd at T =1.4 K obtained by varying the magnetic field at a constant frequency (panel a) and by changing the frequency at a constant magnetic field (panel b). Data have been fit according to a Gaussian profile.}
	\label{spectra}
\end{figure}
A ($\pi/2$ - $\pi$) pulse sequence was employed to collect Cr$_8$Cd NMR spectra with two complementary methods: by varying the frequency at a fixed external magnetic field (frequency-\emph{sweep}, i.e., the frequency is changed in fine steps), or by varying the field in small steps at fixed frequency (field-\emph{sweep}). Signal intensities have been collected point by point as a function of both field and frequency, by integrating the whole spin-echo over time. Our experimental configuration allowed us to detect resonance frequencies in the 17.5-35 MHz range. The detection of $^{53}$Cr resonances below 17 MHz was hindered by the drop of sensitivity ($\propto\nu^2$) and by the dramatic increase of the receiver dead-time at decreasing frequency, mostly arising from spurious magneto-acoustic couplings (\emph{ringing}) of the pick-up coil.
Both field- and frequency-\emph{sweep} $^{53}$Cr-NMR spectra of Cr$_8$Cd are reported in Fig.\ref{spectra}. Resonance peaks have then been fitted with Gaussians line-shapes, in order to extract the resonance frequencies reported as black scatters in Fig.\ref{intensityplot}, where they are also compared with theoretical calculations (more details in the following section). 
\begin{figure}[!h]
	\centering
	\includegraphics[width=0.5\textwidth]{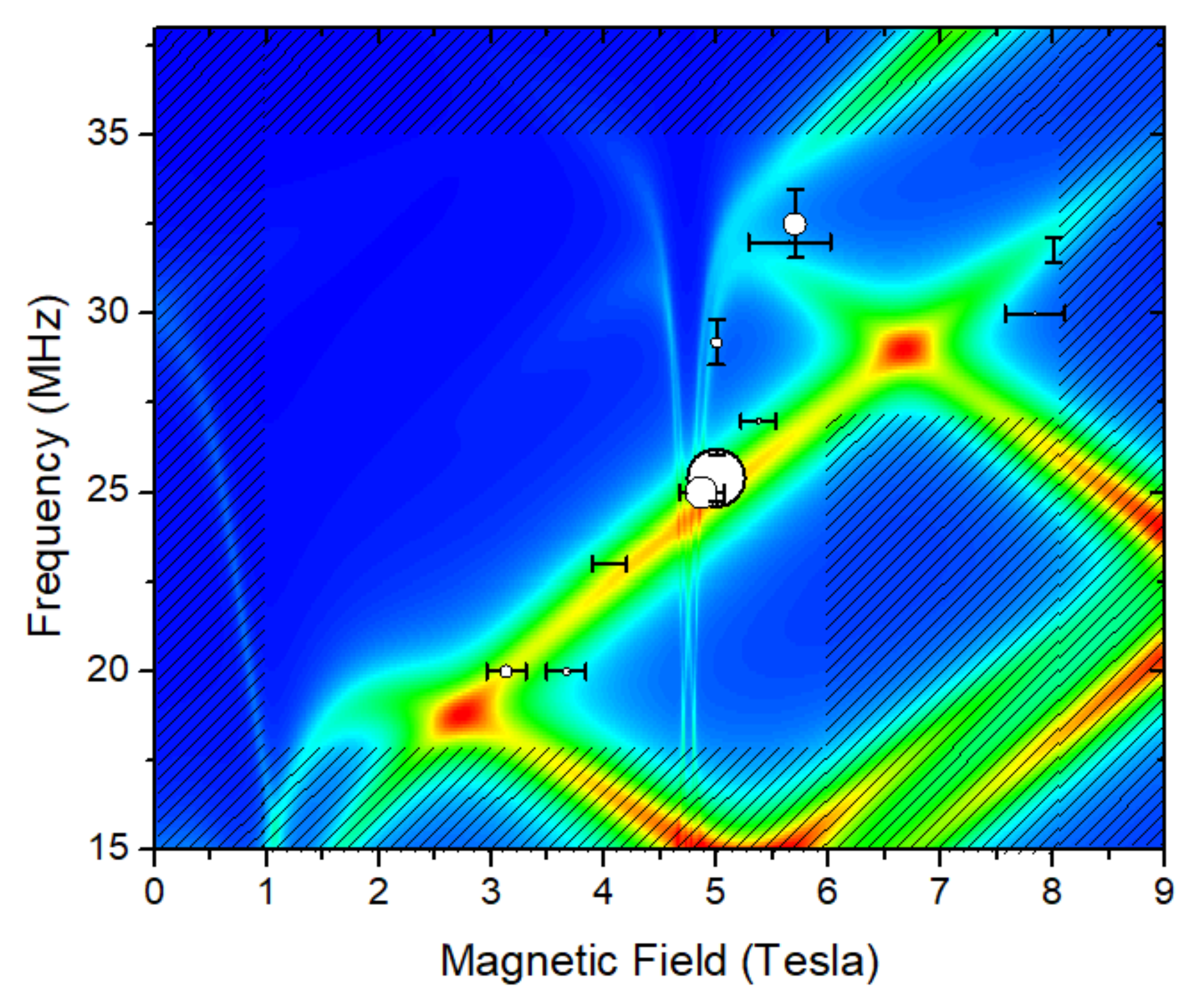}
	\caption{Resonance frequencies calculated with Eq.\ref{contact} as a function of the external magnetic field in the explored frequency range. To each frequency we also associated a Gaussian line-shape with $\sigma$=1 MHz and amplitude given by the level population, yielding the above intensity plot. Black scatters: experimental resonance frequencies obtained by fitting field- and frequency-\emph{sweep} $^{53}$Cr-NMR spectra in Fig.\ref{spectra}. The dimension of the scatters reflect the relative intensities of the detected transitions (in arb. units). Only resonances with a significant intensity are reported, shaded areas have not been experimentally explored. }
	\label{intensityplot}
\end{figure}

\section{Analysis of the data and discussion}

\begin{figure}[!h]
	\centering
	\includegraphics[width=0.48\textwidth]{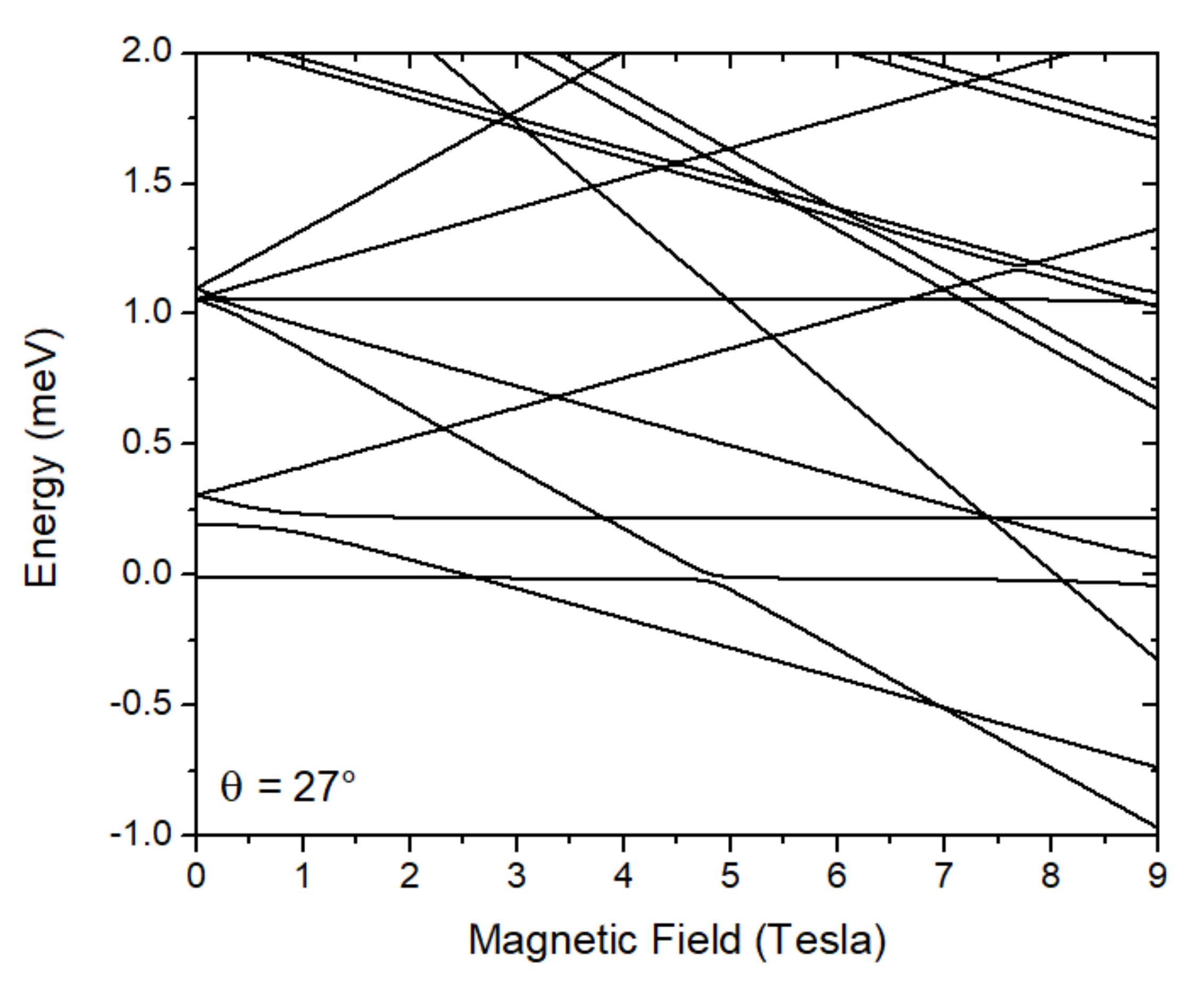}
	\caption{Low-lying energy levels of Cr$_8$Cd as a function of the external magnetic	field calculated for $\theta = {27}^\circ$.}
	\label{levels}
\end{figure}

\begin{figure}[!h]
	\centering
	\includegraphics[width=0.48\textwidth]{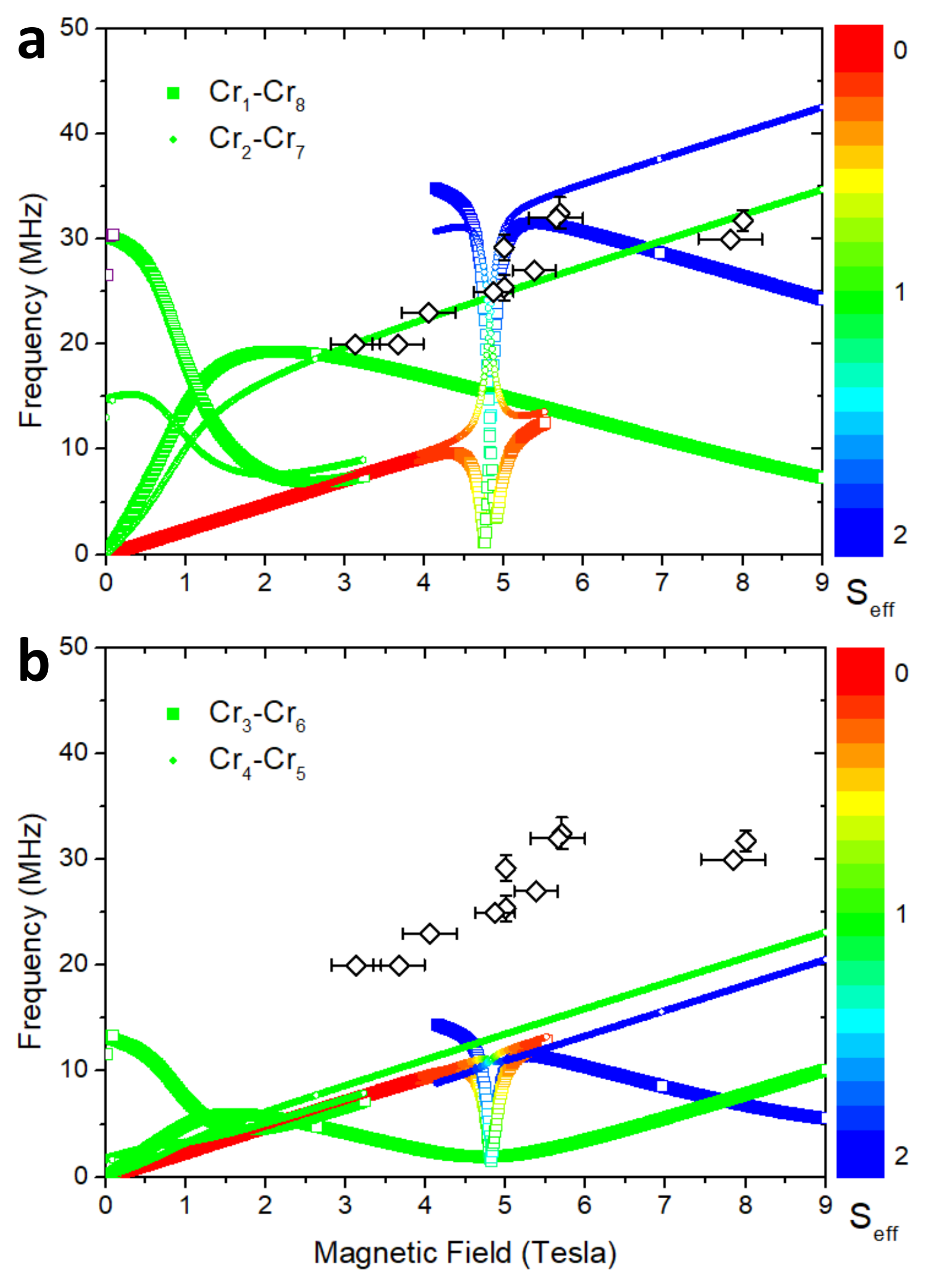}
	\caption{Calculated resonance frequencies for each magnetically-equivalent pair of ions: Cr$_{1}$-Cr$_{8}$ and Cr$_{2}$-Cr$_{7}$ (panel a) and Cr$_{3}$-Cr$_{6}$ and Cr$_{4}$-Cr$_{5}$ (panel b) (see also Fig.\ref{molecule}-b). The colour-map indicates the effective total-spin $S_{\mathrm{eff}}$ (where $\langle S^2\rangle = S_{\mathrm{eff}}(S_{\mathrm{eff}}+1)$) of the populated multiplets considered in the calculation (only levels with a non-negligible population at 1.4 K have been considered). Black scatters represent both field- and frequency-\emph{sweep} experimental data. These figures demonstrate that only $^{53}$Cr nuclei on 1-8 and 2-7 sites have been detected in the explored frequency range. Dashed lines indicate the lower limit of the experimental frequency range.}
	\label{sites}
\end{figure}

The magnetic properties of each Cr$_8$Cd molecule can be described by the Hamiltonian:
\begin{eqnarray}\label{Hamil}
\mathcal{H} &= J\sum_{i=1}^{7}\mathbf{s}_i\cdot\mathbf{s}_{i+1} + d\sum_{i=1}^{8}[s_{z,i}^{2}-s_{i}(s_{i}+1)/3]\\
\nonumber &+\mu_{B}\sum_{i=1}^{8}g\mathbf{s}_{i}\cdot\mathbf{B},
\end{eqnarray}
where $\mathbf{s}_{i}$ is the spin operator of the i$^{th}$ ion in the ring. The first term is the dominant nearest-neighbour isotropic exchange interaction, the second one is the uniaxial single-ion zero-field splitting term (where the $z$ axis is perpendicular to the plane of the ring) and the last term is the Zeeman interaction with an external magnetic field $\mathbf{B}$. Spin Hamiltonian parameters have been previously determined by INS measurements on the parent Cr$_8$Zn compound \cite{Bianchi2009} and magnetization measurements on Cr$_8$Cd \cite{Furukava2008}, yielding $J$=1.32 meV, $d$=0.036 meV and $g$=1.98. Hereinafter, we label Cr$_8$Cd eigenstates by their dominant total-spin $S$. Indeed, even if it is not a good quantum number for Eq.\ref{Hamil}, Cr$_8$Cd low-energy eigenstates are characterised by a very small S-mixing, with dominant component on the total-spin basis of $>99\%$ (except for the very small field ranges around anti-crossings). This result is a consequence of the Heisenberg exchange interaction being the dominant one in the spin Hamiltonian \ref{Hamil}. Our total-spin labelling is also supported by the evaluation of the effective total-spin $S_{\mathrm{eff}}$, where $\langle S^2\rangle = S_{\mathrm{eff}}(S_{\mathrm{eff}}+1)$ (calculations are reported as a colour-map in Fig.\ref{sites}).  
Energy levels as a function of the applied magnetic field with $\theta = {27}^\circ$ are reported in Fig.\ref{levels}. At zero-field Cr$_8$Cd has a non-magnetic $S=0$ ground state, which becomes and $S=1$ after a level crossing at 2.6 T and an $S=2$ after another crossing at 6.9 T. An anti-crossing between the $S = 0$ and the $S=2$ multiplets is also present at 4.8 T.\\
If we take into account the presence of the magnetic $^{53}$Cr nuclei and their interaction with electronic spins, the full Hamiltonian for a Cr$_8$Cd molecule can be written as:
\begin{eqnarray}\label{NHamil}
\mathcal{H} = \mathcal{H}_{el} + \mathcal{H}_{n},
\end{eqnarray}
where $\mathcal{H}_{el}$ is the electronic spin Hamiltonian in Eq.\ref{Hamil} and $\mathcal{H}_{n}$ for the i$^{th}$ Cr ion in the ring is given by:
\begin{eqnarray}\label{NHamil2}
\mathcal{H}_{n,i} = A \mathbf{s}_{i}\cdot\mathbf{I}_{i} + g_{I}\mu_{N}\mathbf{I}_{i}\cdot\mathbf{B}.
\end{eqnarray}
The first term in Eq.\ref{NHamil2} is the hyperfine coupling (assuming an isotropic coupling constant A) while the second one is the nuclear Zeeman term. \footnote{Since we measured only the central line transition ($M_I = 1/2 \leftrightarrow M_I = -1/2$) of the $^{53}$Cr NMR spectrum, we neglect second-order effects due to quadrupolar interactions \cite{Micotti2006,JPCM2012}.}\\
Eigenvectors of the full Hamiltonian in Eq.\ref{NHamil} can be written as product states between the electronic and nuclear eigenfunctions, with well-defined electronic and nuclear quantum numbers. Given its low natural abundance, we can simply assume the presence of one magnetic $^{53}$Cr nucleus per ring. We label the eigenstates as $|\alpha\rangle = |S M_S, M_I\rangle$, where $S$ labels the electronic total-spin multiplet and the quantization axis for $M_S$ and $M_I$ coincides with the direction of the static magnetic field. We have also checked that electronic spins are practically aligned with the direction of the applied field already at the lowest explored magnetic fields ($>$ 1 T). Since our NMR experiment was performed at T = 1.4 K, it is reasonable to consider nuclear transitions between hyperfine levels within the lowest-energy total-spin multiplets. 
\begin{figure}[!h]
	\centering
	\includegraphics[width=0.48\textwidth]{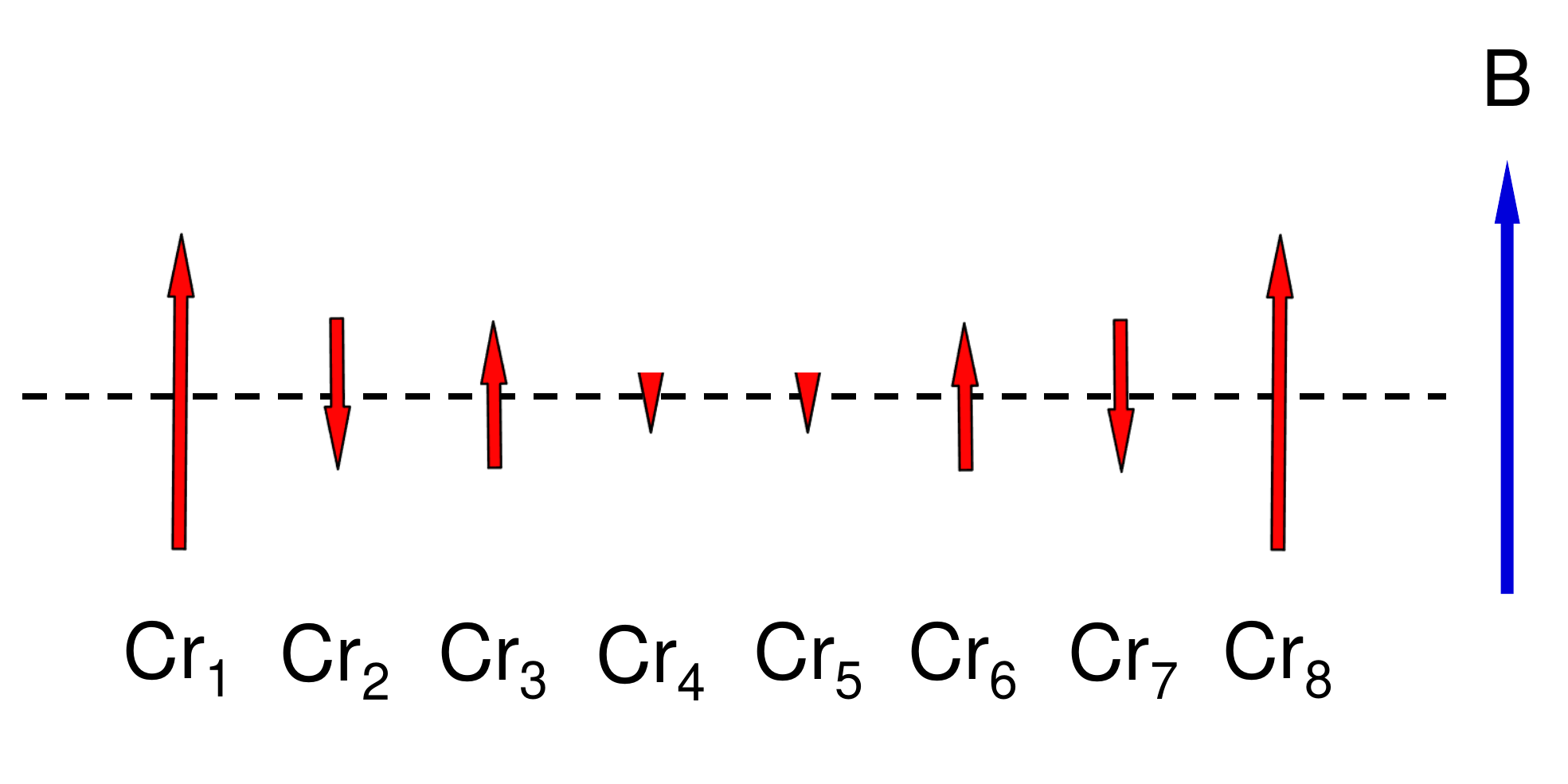}
	\caption{Schematic representation of the NC spin arrangement of the Cr$_8$Cd ring. Red arrows represent the calculated expectation values of the local spins along the direction of the static magnetic field z' at 5 T (when the ground state is an $S = 1$).}
	\label{NC}
\end{figure}\\
Resonance frequencies can be expressed in terms of the Larmor frequency of $^{53}$Cr, shifted by the isotropic core polarization hyperfine contact term \cite{Fermi}:
\begin{equation}\label{contact}
\nu_{\alpha,\alpha'}^{i} = \frac{\gamma}{2\pi}(B_{z'} - g A_C \langle s_{z',i} \rangle_{S,M_S}),
\end{equation}
where $z'$ is the direction of the static magnetic field, $\gamma = g_I\mu_N$ is the $^{53}$Cr gyromagnetic ratio, $A_C$ is the core polarization field, the only free parameter of our model for Cr$_8$Cd, and $\langle s_{z',i} \rangle_{S,M_S} = \langle S M_S|s_{z',i}|S M_S\rangle$ is the expectation value of the local spin operator of the i$^{th}$ ion in the ring \cite{Micotti2006,JPCM2012}.\\ 
The absorptive part of the dynamic susceptibility for the nuclear transitions between two levels $|\alpha\rangle = |S M_S, M_I\rangle$ and $|\alpha'\rangle = |S M_S, M_{I\pm1}\rangle$ can be written as \cite{Jensen}:
\begin{eqnarray}\label{chi}
\chi''(\omega) &\propto \sum_{\alpha,\alpha'}\frac{n_{\alpha}\beta(E_{\alpha'} - E_{\alpha})}{\Gamma_{\alpha,\alpha'}\sqrt{2\pi}}\times\\
\nonumber 	   &\times\exp(\frac{-(\hbar\omega - (E_{\alpha'} - E_{\alpha})^2}{2\Gamma_{\alpha,\alpha'}})\times\\
\nonumber &\times\{\langle\alpha|g_{I}\mu_{N}I_{x'}|\alpha'\rangle\langle\alpha'|g_{I}\mu_{N}I_{x'}|\alpha\rangle+\\ \nonumber&+\langle\alpha|g_{I}\mu_{N}I_{y'}|\alpha'\rangle\langle\alpha'|g_{I}\mu_{N}I_{y'}|\alpha\rangle\}.
\end{eqnarray}
$n_{\alpha}$ is the thermal population of the level $\alpha$, $\beta=1/K_BT$ and $(E_{\alpha'} - E_{\alpha})=\pm(Am_S + g_I\mu_NB_{z'})=\hbar\omega_{\alpha,\alpha'}$, where $m_S$ is the local spin operator expectation value $\langle s_{z',i} \rangle_{S,M_S}$. At last, $\Gamma_{\alpha,\alpha'}$ is the width of the Gaussian line-shape. Thus, for nuclear transitions between hyperfine levels within the low-energy total-spin multiplets of Cr$_8$Cd we do expect resonance-frequency shifts to depend on the expectation value of the local spin operator $\langle s_{z',i} \rangle$ and intensities to depend on the thermal population $n_{\alpha}$, which is mostly set by the electronic component\footnote{The difference $(E_{\alpha'} - E_{\alpha})$ in Eq.\ref{chi} is weakly field dependent, because the main contribution comes from hyperfine interaction, and it is small with respect to $\beta=1/K_BT$.}.\\
In order to determine the core polarization field $A_C$ for Cr$_8$Cd, we calculated the resonance frequencies with Eq.\ref{contact} as a function of the applied static magnetic field. Expectation values of the local spin operators $\langle s_{z',i} \rangle_{S,M_S}$  were calculated on each considered eigenstate by diagonalizing the spin Hamiltonian in Eq.\ref{Hamil}. Fig.\ref{NC} reports the calculated expectation values for a static magnetic field of 5 T on the magnetic ground state, i.e. when it is a magnetic $S = 1$. The spins are arranged in a NC configuration. We also simulated the intensities of each transition, assuming a Gaussian line-shape with $\sigma$=1 MHz and amplitude given by the population $n_{\alpha}$. Theoretical results are reported in the intensity plot in Fig.\ref{intensityplot}, compared with experimental frequencies (black scatters). The latter can be reproduced by assuming a core polarization field $A_C$ = -12.7 T/$\mu_B$, in agreement with the results on the parent compounds Cr$_7$Cd and Cr$_7$Ni (see Table \ref{table2} and Ref.\cite{Micotti2006,JPCM2012}).
\begin{table}[!h]
	\centering
	
	\begin{tabular}{|l|l|}
		\hline
		 AF ring & A$_C$ (T/$\mu_B$)  \\
		\hline
		Cr$_8$Cd       & -12.7 	\\
		\hline
		Cr$_7$Cd       & -12.38 	\\
		\hline
		Cr$_7$Ni       & -11 	\\
		\hline
	\end{tabular}
	\caption{Comparison of core polarization constant A$_C$ obtained with $^{53}$Cr-NMR experiments on Cr$_8$Cd (present work),  Cr$_7$Cd \cite{Micotti2006} and Cr$_7$Ni \cite{JPCM2012}.} \label{table2}
\end{table}\\
The next step was to attribute the detected signals to the corresponding $^{53}$Cr nuclei along the ring. Fig.\ref{sites}-a shows the calculated frequencies for the Cr$_{1}$-Cr$_{8}$ and Cr$_{2}$-Cr$_{7}$ magnetically-equivalent sites (see Fig.\ref{molecule}-b), compared with the experimental data. It is evident that the signals observed can be totally ascribed to these Cr nuclei. Given the theoretically predicted NC configuration (see Fig.\ref{NC}), Cr$_{1}$-Cr$_{8}$ at the extremities of the chain have a significant electronic magnetic moment along the direction of the applied field, inducing a high resonance-frequency shift. Their nearest-neighbours Cr$_{2}$-Cr$_{7}$ have a smaller electronic moment, but anti-parallel to the applied field. Thus, given $A_C < 0$, the two contributions to Eq.\ref{contact} add up, yielding an increased resonance frequency. Cr$_{4}$-Cr$_{5}$ ions, localized in the middle of the ring, are expected to have a negligible magnetic moment along the direction of the applied field, thus inducing a small resonance-frequency shift. The magnetic moment of Cr$_{3}$-Cr$_{6}$ ions is instead comparable with the Cr$_{2}$-Cr$_{7}$ one. However, electronic spins on these sites are parallel to the applied field and thus the two contributions in Eq.\ref{contact} are opposite ($A_C < 0$), yielding small resonance frequencies. In fact, signals due to $^{53}$Cr nuclei on 4-5 and 3-6 sites have not been observed in the spectra. A core polarization field twice as high as -12.7 T/$\mu_B$ would be necessary to shift them in the experimental frequency window (see Fig.\ref{sites}-b), a value not compatible with previous results on similar compounds (see Table \ref{table2}). Therefore, $^{53}$Cr nuclei on 1-8 and 2-7 sites are the only ones that can be detected in the explored frequency range. Thus, $^{53}$Cr-NMR experimental data confirm the NC configuration of the electronic local spin moments along the Cr$_8$Cd ring. At last, Fig.\ref{sites}-a show that the measured signals are mainly due to nuclei whose electronic spins are in has an effective total-spin $S \geq 1$. 

\section{Conclusions}

In conclusion, we have studied the local spin densities along the Cr$_8$Cd AF ring with $^{53}$Cr-NMR. This magnetically-open ring is a model system to study open AF chains with an even number of spins. From the NMR spectra we have extracted low-temperature resonant frequencies of $^{53}$Cr-NMR at different applied magnetic fields. By comparing these results with our calculations, we have obtained a core polarization field for Cr$_8$Cd of $A_C$ = -12.7 T/$\mu_B$, in agreement with the results on the parent compounds Cr$_7$Cd and Cr$_7$Ni. Moreover, we have investigated the expectation values of the local spins $\langle s_{z',i} \rangle$ along the direction of the applied magnetic field, confirming a NC spin configuration, when Cr$_8$Cd is in a magnetic state.\\
We have also demonstrated that with $^{53}$Cr-NMR it is possible to obtain the same information on the spin arrangement of AF rings, which can be extracted from more resource- and time consuming techniques like polarised neutron diffraction. Even if both experiments require suitably large single crystal samples, the flexibility of $^{53}$Cr-NMR and the straightforward data analysis allows one to perform these kind of experiments on rings/chains with an higher number of spins or on more complex systems. In order to characterise new compounds with $^{53}$Cr-NMR more data should be collected with respect to those presented here for the well-known Cr$_8$Cd ring. For instance, more information could be obtained by simply measuring NMR spectra for different directions of the applied magnetic field. Finally, it is worth underlining that NMR is a very effective technique to characterise molecular nanomagnets in view of their application in quantum information processing. Indeed, other experiments can be combined to investigate local spin moments on the magnetic ions exploiting the same NMR apparatus. For instance, measurements of the $^1$H spin-lattice relaxation time T$_1$ can be exploited to probe electronic relaxation as a function of field and temperature \cite{Cr7NiNMR,Fe7,Garlatti2016}. Most importantly, Rabi oscillations of the nuclear magnetization of a $^{173}$Yb qudit has been recently measured with $^{173}$Yb-NMR\cite{Hussain2018}, demonstrating the capability to coherently manipulate qubits and qudits with NMR radio frequencies.

\ack
Authors gratefully acknowledge financial support from FIRB Project No. RBFR12RPD1 and PRIN Project 2015 No. HYFSRT of the Italian MIUR and from the European project QuantERA 2017 - SUMO.

\section*{References}
\bibliography{Cr8Cd}

\providecommand{\newblock}{}
\begin{thebibliography}{10}
\expandafter\ifx\csname url\endcsname\relax
  \def\url#1{{\tt #1}}\fi
\expandafter\ifx\csname urlprefix\endcsname\relax\def\urlprefix{URL }\fi
\providecommand{\eprint}[2][]{\url{#2}}

\bibitem{due}
Haldane F~D~M 1983 {\em Phys. Rev. Lett.\/} {\bf 50} 1153

\bibitem{tre}
Affleck I, Kennedy T and nad H~Tasaki E~H~L 1987 {\em Phys. Rev. Lett.\/} {\bf
  59} 799

\bibitem{edge1}
Hagiwara M, Katsumata K, Affleck I, Halperin B and Renard J 1990 {\em Phys.
  Rev. Lett.\/} {\bf 65} 3181

\bibitem{dodici}
Miyashita S and Yamamoto S 1993 {\em Phys. Rev. B\/} {\bf 48} 913

\bibitem{edge3}
Qin S, Ng T~K and Su Z~B 1995 {\em Phys. Rev. B\/} {\bf 57} 12844

\bibitem{undici}
Lounis S, Dederichs P and Blügel S 2008 {\em Phys. Rev. Lett.\/} {\bf 101}
  107204

\bibitem{Politi}
Politi P and Pini M 2009 {\em Phys. Rev. B\/} {\bf 79} 012405

\bibitem{tredici}
Machens A, nad O~Waldmann N~P~K, Schneider I and Eggert S 2013 {\em Phys. Rev.
  B\/} {\bf 87} 144409

\bibitem{otto}
Spinelli A, Bryant B, adn J~Fernandez-Rossier F~D and Otte A~F 2014 {\em Nat.
  Mater.\/} {\bf 13} 782

\bibitem{nove}
Khajetoorians A~A, Wiebe J, Chilian B, Lounis S, Blügel S and Wiesendanger R
  2012 {\em Nat. Phys.\/} {\bf 8} 497

\bibitem{dieci}
Loth S, Baumann S, Lutz C, Eigler D and Heinrich A 2012 {\em Science\/} {\bf
  335} 196

\bibitem{Sessoli1993}
Sessoli R, Gatteschi D, Caneschi A and Novak M~A 1993 {\em Nature\/} {\bf 365}
  141

\bibitem{Carretta2003}
Carretta S, van Slageren J, T~Guidi E~L, Mondelli C, Rovai D, Cornia A, Dearden
  A~L, Carsughi F, Affronte M, Frost C~D, Winpenny R~E~P, Gatteschi D, Amoretti
  G and Caciuffo R 2003 {\em Phys. Rev. B\/} {\bf 67} 094405

\bibitem{Affronte2003}
Affronte M, Guidi T, Caciuffo R, Carretta S, Amoretti G, Hinderer J, Sheikin I,
  Jansen A~G~M, Smith A~A, Winpenny R~E~P, van Slageren J and Gatteschi D 2003
  {\em Phys. Rev. B\/} {\bf 68} 104403

\bibitem{Waldmann2003}
Waldmann O, Guidi T, Carretta S, Mondelli C and Dearden A~L 2003 {\em Phys.
  Rev. Lett.\/} {\bf 91} 23720

\bibitem{NatPhys}
Baker M~L, Guidi T, Carretta S, Ollivier J, Mutka H, Gudel H~U, Timco G~A,
  McInnes E~J~L, Amoretti G, PWinpenny R~E and Santini P 2012 {\em Nat.
  Phys.\/} {\bf 8} 906

\bibitem{Baker2012}
Baker M~L, Timco G~A, Piligkos S, Mathieson J~S, Mutka H, F~Tuna P~K, Antkowiak
  M, T~Guidi T~Gupta H~R, JWoolfson R, amd R~G~Pritchard G~K, HWeihe, Cronin L,
  Rajaraman G, Collison D, McInnes E~J~L and Winpenny R~E~P 2012 {\em PNAS\/}
  {\bf 109} 19113

\bibitem{Waldmann2009}
Waldmann O, Stamatatos T~C, Christou G, H~U~G¨udel I~S and Mutka H 2009 {\em
  Phys. Rev. Lett.\/} {\bf 102} 157202

\bibitem{Ummethum2012}
Ummethum J, Nehrkorn J, Mukherjee S, Ivanov N~B, S~Stuiber T~Str¨assle
  P~L~W~T~P, Mutka H, GChristou, Waldmann O and Schnack J 2012 {\em Phys. Rev.
  B\/} {\bf 86} 104403

\bibitem{Cu3}
Cage B, Cotton F~A, Dalal N~S, Hillard E~A, Rakvin B and Ramsey C~M 2003 {\em
  J. Am. Chem. Soc.\/} {\bf 125} 5270

\bibitem{Cr8Ni1}
Cador O, Gatteschi D, Sessoli R, Larsen F~K, Overgaard J, ABarra, Teat S~J,
  Timco G~A and Winpenny R 2004 {\em Angew. Chem. Int. Ed.\/} {\bf 43} 5196

\bibitem{Cr8Ni2}
Furukawa Y, Kiuchi K, Kumagai K~I, Ajiro Y, Y~Narumi M~I, Kindo K, Bianchi A,
  Carretta S, Santini P, Borsa F, Timco G~A and Winpenny R~E~P 2009 {\em Phys.
  Rev. B\/} {\bf 79} 134416

\bibitem{VO7}
Hoshino N, Nakano M, Nojiri H, Wernsdorfer W and Oshio H 2009 {\em J. Am. Chem.
  Soc.\/} {\bf 131} 15100

\bibitem{Antkowiak2013}
Antkowiak M, Kozlowski P, Kamieniarz G, Timco G~A, Tuna F and Winpenny R~E~P
  2013 {\em Phys. Rev. B\/} {\bf 87} 184430

\bibitem{Kamieniarz2015}
Kamieniarz G, Florek W and Antkowiak M 2015 {\em Phys. Rev. B\/} {\bf 92}
  140411(R)

\bibitem{Garlatti2016}
Garlatti E, Bordignon S, Carretta S, Allodi G, Amoretti G, DeRenzi R,
  Lascialfari A, Furukawa Y, Timco G~A, Woolfson R, Winpenny R~E~P and Santini
  P 2016 {\em Phys. Rev. B\/} {\bf 93} 024424

\bibitem{Micotti2006}
Micotti E, Furukawa Y, Kumagai K, Carretta S, Lascialfari A, Borsa F, Timco G~A
  and PWinpenny R~E 2006 {\em Phys. Rev. Lett.\/} {\bf 97} 267204

\bibitem{Piligkos2007}
Piligkos S, Bill E, Collison D, McInnes E~J~L, Timco G~A, Weihe H, Winpenny
  R~E~P and Neese F 2007 {\em J. Am. Chem. Soc.\/} {\bf 129} 760

\bibitem{Timco2005}
Timco G~A, Batsanov A~S, Larsen F~K, Muryn C~A, Overgaard J, Teate S~J and
  Winpenny R~E~P 2005 {\em Chem. Commun.\/}  3649--3651

\bibitem{Furukava2008}
Furukawa Y, Kiuchi K, ichi Kumagaia K, Ajiro Y, Narumi Y, Iwaki M, Kindo K,
  Bianchi A, Carretta S, Timco G~A and Winpenny R~E~P 2008 {\em Phys.Rev. B\/}
  {\bf 78} 092402

\bibitem{Bianchi2009}
Bianchi A, Carretta S, Santini P, Amoretti G, Guidi T, Qiu Y, Copley J~R~D,
  Timco G, Muryn A and Winpenny R~E~P 2009 {\em Phys.Rev. B\/} {\bf 79} 144422

\bibitem{Adelnia2015}
Adelnia F, Chiesa A, Bordignon S, Carretta S, Ghirri A, Candini A, Cervetti C,
  Evangelisti M, Affronte M, Sheikin I, Winpenny R, Timco G, Borsa F and
  Lascialfari A 2015 {\em J. Chem. Phys.\/} {\bf 143} 244321

\bibitem{Larsen2003}
Larsen F~K, McInnes E~J~L, Mkami H~E, Overgaard J, Piligkos S, Rajaraman G,
  Rentschler E, Smith A~A, Smith G~M, Boote V, Jennings M, Timco G~A and
  Winpenny R~E~P 2003 {\em Angew. Chem.\/} {\bf 115} 105

\bibitem{Guidi2005}
Caciuffo R, Guidi T, Amoretti G, Carretta S, Liviotti E, Santini P, Mondelli C,
  Timco G, Muryn C~A and Winpenny R~E~P 2005 {\em Phys. Rev. B\/} {\bf 71}
  174407

\bibitem{Garlatti2014b}
Garlatti E, Albring M~A, Baker M~L, Docherty R~J, Mutka H, Guidi T, Sakai V~G,
  Whitehead G~F~S, Pritchard R~G, Timco G~A, Tuna F, Amoretti G, Carretta S,
  Santini P, Lorusso G, Affronte M, McInnes E~J~L, Collison D and Winpenny
  R~E~P 2014 {\em J. Am. Chem. Soc.\/} {\bf 136} 9763

\bibitem{Cr7Co}
Garlatti E, Guidi T, Chiesa A, Ansbro S, Baker M~L, Ollivier J, Mutka H, Timco
  G~A, Vitorica-Yrezabal I, Pavarini E, Santini P, Amoretti G, Winpenny R~E~P
  and Carretta S 2018 {\em Chem. Sci.\/} {\bf 9} 3555

\bibitem{Troiani20051}
Troiani F, Affronte M, Carretta S, Santini P and Amoretti G 2005 {\em Phys.Rev.
  Lett.\/} {\bf 94} 190501

\bibitem{Troiani20052}
Troiani F, Ghirri A, Affronte M, Carretta S, Santini P, Amoretti G, S~Piligkos
  G~T and Winpenny R~E~P 2005 {\em Phys.Rev. Lett.\/} {\bf 94} 207208

\bibitem{NatNano}
Timco G~A, Carretta S, Troiani F, Tuna F, Pritchard R~J, AMuryn C, McInnes
  E~J~L, Ghirri A, Candini A, Santini P, Amoretti G, Affronte M and PWinpenny
  R~E 2009 {\em Nat. Nanotech.\/} {\bf 4} 173

\bibitem{Wedge2012}
JWedge C, Timco G, Spielberg E~T, George R~E, Tuna F, Rigby S, McInnes E~J~L,
  Winpenny R~E~P, Blundell S~J and Ardavan A 2012 {\em Phys. Rev. Lett.\/} {\bf
  108} 107204

\bibitem{Ferrando2016}
Ferrando-Soria J, Moreno-Pineda E, Chiesa A, Fernandez A, Magee S, Carretta S,
  Santini P, Vitorica-Yrzebal I, Tuna F, Timco G~A, McInnes E~J~L and Winpenny
  R~P 2016 {\em Nat. Commun.\/} {\bf 7} 11377

\bibitem{Garlatti2017}
Garlatti E, Guidi T, Ansbro S, Santini P, Amoretti G, Ollivier J, Mutka H,
  Timco G, Vitorica-Yrezabal I, Whitehead G, Winpenny R and Carretta S 2017
  {\em Nat. Commun.\/} {\bf 8} 14543

\bibitem{JPCM2012}
Casadei C~M, Bordonali L, Furukawa Y, Borsa F, Garlatti E, Lascialfari A,
  Carretta S, Sanna S, Timco G and Winpenny R 2012 {\em J. Phys.: Cond.
  Matter\/} {\bf 24} 406002

\bibitem{NatCommun2015}
Guidi T, Gillon B, Mason S, Garlatti E, Carretta S, Santini P, Stunault A,
  Caciuffo R, van Slageren J, Klemke B, Cousson A, Timco G and Winpenny R 2015
  {\em Nat. Commun.\/} {\bf 6} 7061

\bibitem{Hyrespect}
Allodi G, Banderini A, Renzi R~D and Vignali C 2005 {\em Rev. Sci. Instrum.\/}
  {\bf 76} 1

\bibitem{Fermi}
Fermi E 1930 {\em Z. Phys.\/} {\bf 60} 320

\bibitem{Jensen}
Jensen J and Mackintosh A~K 1991 {\em Rare-Earths Magnetism\/} (Oxford:
  Clarendon Press)

\bibitem{Cr7NiNMR}
Bianchi A, Carretta S, Santini P, Amoretti G, Lago J, Corti M, Lascialfari A,
  Arosio P, Timco G and Winpenny R~E~P 2010 {\em Phys. Rev. B\/} {\bf 82}
  134403

\bibitem{Fe7}
Garlatti E, Carretta S, Santini P, Amoretti G, Mariani M, Lascialfari A, Sanna
  S, Mason K, Chang J, Tasker P and Brechin E~K 2013 {\em Phys. Rev. B\/} {\bf
  87} 054409

\bibitem{Hussain2018}
Hussain R, Allodi G, Chiesa A, Garlatti E, Mitcov D, Konstantatos A, Pedersen
  K~S, Renzi R~D, Piligkos S and Carretta S 2018 {\em J. Am. Chem. Soc.\/} {\bf
  140} 9814

\end{thebibliography}

\end{document}